\documentclass{Rinton-P9x6}
\input epsf.tex
\newcommand{\mpt}{\mbox{${p\!\!\!/_T}$}}

\def\DESepsf(#1 width #2){\epsfxsize=#2 \epsfbox{#1}}
\begin{document}

\title{mSUGRA At A 500-GeV Linear Collider}

\author{R. Arnowitt}

\address{Center for Theoretical Physics, Department of 
Physics, Texas A\&M University, College Station TX 77843-4242, USA}
\author{B. Dutta}

\address{Department of Physics, 
University of Regina, Regina SK, S4S OA2, Canada}
\author{T. Kamon and V. Khotilovich}

\address{Department of Physics, Texas 
A\&M University, College Station TX, 77843-4242, USA}


\maketitle

\abstracts{A study is made of what SUSY signals would be observable for mSUGRA models 
in a 500-GeV linear collider. All current experimental bounds on the mSUGRA 
parameter space are imposed. For $m_0  <$ 1 TeV (or alternately if the current 
$g_\mu - 2$ anomaly maintains) the only observable signals that remain are 
slepton pair production and neutralino production of 
${\tilde\chi^0_2}+{\tilde\chi^0_1}$. 
Slepton pair production can occur for masses $<$ 250 GeV which for the 
selectron and smuon pairs require $\tan\beta <$ 40. In this domain very 
accurate selectron and smuon masses could be measured. Light staus, 
$\tilde\tau_1$, 
with mass $<$ 250 GeV can be pair produced for any $\tan\beta$ and the neutralino 
signal can be seen provided $m_{1/2}\stackrel{<}{\sim}400$ GeV. However, the detection of 
these requires a much more complicated analysis due to the fact that the 
dark matter co-annihilation constraint requires that the $\tilde\tau_1$ and 
${\tilde\chi^0_1}$ 
mass difference be $\stackrel{<}{\sim}$ 15 GeV. The point $m_{1/2}$ = 360 GeV,
 $A_0 = 0$, $\mu>0$ is 
analyzed in detail, and it is shown that the stau and neutralino signals 
can be detected provided an active mask down to $2^o$ is used. However, large 
parts of the mSUGRA parameter space exists where a 500-GeV machine would 
not be able to see any SUSY signal.
}
\section{Introduction}

There is a growing consensus that the next high energy machine to be built 
after the Large Hadron Collider (LHC) should be an electron-positron linear 
collider (LC), and initial designs call for a 500-GeV machine. While the 
technology to be chosen (TESLA or NLC/JLC) and where the siting should be 
is still under discussion, the proponents believe that such a machine is 
technically feasible. We consider here what aspects of mSUGRA might be 
tested at a 500-GeV LC.

There has been in the past a huge amount of analysis on methods of 
detecting SUSY at LCs. However, the minimal supergravity 
model, mSUGRA \cite{sugra1,sugra2,nilles},  has several special aspects that make its 
predictions clearer and hence more directly accessible to experimental 
study. Hence it is worthwhile to examine this particular model. Thus:

\noindent(1) mSUGRA depends on only four additional parameters and one sign beyond those of the 
Standard Model (SM). These are $m_0$ (the universal soft breaking mass at the 
GUT scale $M_G$); $m_{1/2}$ (the universal gaugino soft breaking mass at $M_G$) ; 
$A_0$ (the universal cubic soft breaking mass at $M_G$); $\tan\beta = <H_2>/<H_1>$ 
at the  electroweak scale (where $H_2$ gives rise to u quark masses and $H_1$ 
to d quark and lepton masses); and the sign of $\mu$, the Higgs mixing 
parameter in the superpotential ($W_{\mu} = \mu H_1 H_2$). Note that the lightest 
neutralino $\tilde\chi^0_1$ and the gluino $\tilde g$ are approximately related to 
$m_{1/2}$ by 
$m_{\tilde\chi^0_1}\cong 0.4m_{1/2}$ and $m_{\tilde g}\cong2.8m_{1/2}$ . We will examine here the following 
parameter range: 
$0<m_{1/2} < 1$TeV,  
$|A_0| <  4 m_{1/2}$ and $1 < \tan\beta < 55$. The 
$m_{1/2}$ range thus covers the limit of gluino discovery at the LHC.

\noindent(2) mSUGRA makes predictions for a wide range of phenomena, and thus the 
model is already significantly constrained by experiment. Most important 
for limiting the parameter space are:\\
(i)   The light Higgs mass bound from LEP \cite{higgs1} $m_h > 114$ GeV . Since 
theoretical calculations of $m_h$ still have a 2-3 GeV error, we will 
conservatively assume this to mean that $(m_h)^{\rm theory} > 111$ GeV.
(ii)  The $b\rightarrow s + \gamma$ branching ratio\cite{bsgamma}. We assume here a relatively 
broad range (since there are theoretical errors in extracting the branching 
ratio from the data):

\begin{equation} 1.8\times10^{-4} < B(B \rightarrow X_s \gamma) <
4.5\times10^{-4}
\label{bs}
\end{equation}

\noindent(iii) In mSUGRA the lightest neutralino, $\tilde\chi^0_1$, is the candidate for dark 
matter (DM). Previous bounds  from balloon flights (Boomerang, Maxima, 
Dasi, etc.) gave a relic density bound for DM of $0.07 < \Omega_{DM} h^2 < 0.21 $
(where $\Omega_{DM}$ is the density of dark matter relative to the critical 
density to close the universe, and $h = H/100km/sec Mpc$ where $H$ is the Hubble 
constant). However, the new data from WMAP \cite{sp} greatly tightens this (by a 
factor of four) and the 2$\sigma$ bound is now:
\begin{equation}
 0.095 < \Omega_{\rm DM} h^2 <0.129
\label{om}\end{equation}

\noindent(iv)  The bound on the lightest chargino mass \cite{aleph}: $\tilde\chi^\pm_1>$ 104 GeV

\noindent(v)   The muon magnetic moment anomaly, $\delta a_\mu$ \cite{BNL}. Here the 
calculation of the leading order of the hadronic SM contribution is still
in doubt. Using the $e^+$$e^-$ data to calculate this, one gets a 3$\sigma$
deviation of the SM from the experimental result \cite{dav,hag,jeg}, while using tau 
decay and CVC  analysis with CVC breaking included one get a 1$\sigma$ 
deviation \cite{dav}. However, comparison between the $e^+$$e^-$ analysis and the tau 
decay analysis exhibits more than a 4 $\sigma$ disagreement in one channel making it 
difficult to argue one should average the two results. Most recently \cite{akh} CMD-2 has done a reanalysis
of their  $e^+$$e^-$ data correcting their treatment of vacuum polarization diagrams. This reduces
the  $e^+$$e^-$ disagreement with the SM to perhaps 2$\sigma$. Thus the 
situation is still much up in the air, and future data from KLOE, BaBar and 
Belle as well as additional BNL E821 data for the $\mu^-$ may help to clarify matters in 
the future.

\noindent(3) One can now qualitatively state the constraints on the parameter space 
produced by the above experimental bounds:

\noindent(i)   The relic density constraint produces a narrow rising band of allowed 
parameter space in the $m_0$ - $m_{1/2}$ plane.

\noindent(ii)   In this band, the $m_h$ and $b\rightarrow s+\gamma$ constraints produce a lower 
bound on $m_{1/2}$ for all $\tan\beta$:

\begin{equation}
m_{1/2}\stackrel{>}{\sim}  300 
{\rm GeV}  
\label{mhalf}\end{equation}                                        
which implies $m_{\tilde\chi^0_1}> 120$ GeV and $m_{\tilde g} >$ 250 GeV.

\noindent(iii)   If the $g_\mu -2$ effect is real, then $\mu >0$, and the combined effects 
of $g_\mu -2$ and the $M_{\tilde\chi^\pm_1} >$ 104 GeV eliminates all other possible domains 
satisfying the relic density constraint.

In the following, we will analyze the case where $\mu > 0$, but leave open the 
question of the validity of the $g_\mu - 2$ effect. (See also \cite{ellis} for $\mu <0$.) 
Figs. 1, 2 and 3 illustrate the above constraints on the mSUGRA parameter 
space for $\tan\beta$ = 10, 40 and 50  with $A_0$ = 0. The red area is the 
parameter space allowed by the earlier balloon CMB experiments, while the 
(reduced) blue area is the region now allowed by WMAP, Eq. (2). The dotted 
red lines are for different Higgs masses, and the light blue region would 
be excluded if $\delta a_\mu > 11\times10^{-10}$. It is important to note that the 
narrowness of the allowed dark matter band is not a fine tuning. The lower 
limit of the band comes from the rapid annihilation of neutralinos in the 
early universe due to co-annihilation effects as the light stau mass, 
$m_{\tilde\tau_1}$, approaches the neutralino mass as one lowers $m_0$. Thus the lower 
edge of the band corresponds to the lower bound of Eq.(2), and the band is 
cut off sharply due to the Boltzman exponential behavior. The upper limit 
of the band [corresponding to the upper bound of Eq.(2)] arises due to 
insufficient annihilation as $m_0$ is raised. As the WMAP data becomes more 
accurate, the the allowed band will narrow even more. (Note that the 
slope and position of the band changes, however as $A_0$ is changed.) Thus the 
astronomical determination of the amount of dark matter  effectively 
determines one of the four parameters of mSUGRA.

\noindent(4) In order to carry out the calculations it is necessary to include a 
number of corrections to obtain results of sufficient accuracy, and 
we  list some of these here:\\
\noindent(i)   Two loop gauge and one loop Yukawa RGE equations are used from $M_G$ to 
the electroweak scale , and QCD RGE below for the light quarks. Two loop 
and pole mass corrections are included in the calculation of $m_h$.\\
\noindent(ii)   One loop corrections to $m_b$ and $m_\tau$  are included \cite{rattazi}\\
\noindent(iii)  Large $\tan\beta$ SUSY corrections to $b\rightarrow s + \gamma$ are included
\cite{degrassi}\\
\noindent(iv)   All  $\tilde\tau_1$ - $\tilde\chi^0_1$ co-annihilation channels are included in the 
relic density calculation \cite{bdutta}.\\
\noindent We do not include Yukawa unification or proton decay constraints as these 
depend sensitively on post GUT physics, about which little is known.

 \begin{figure}\vspace{-0cm}
\centerline{ \DESepsf(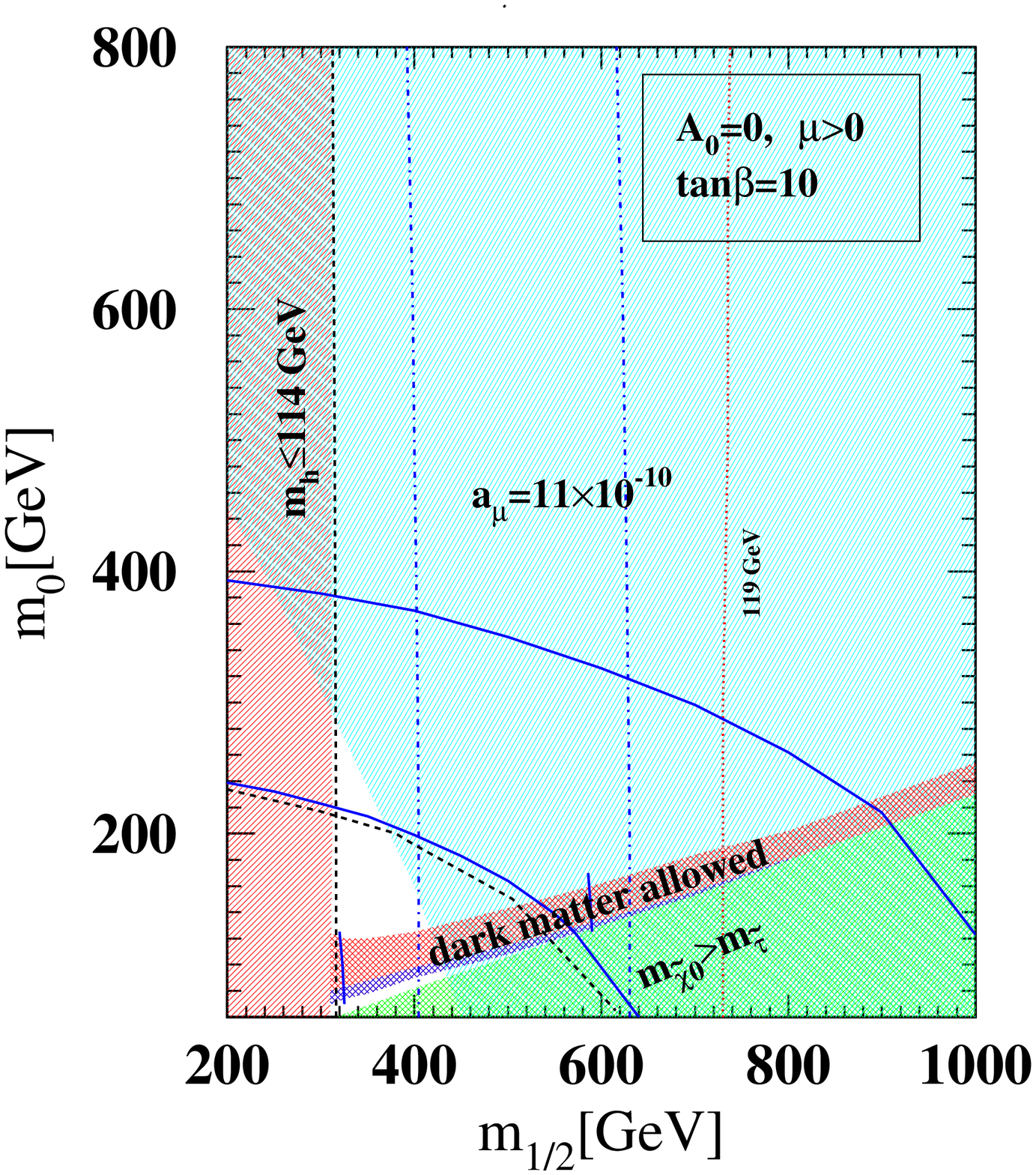 width 8 cm) }
\caption {\label{fig1}  Allowed region in the $m_0$ - $m_{1/2}$ plane from the relic density 
constraint  for $\tan\beta = 10$, $A_0  = 0$ and $\mu >0$. The red region was 
allowed by the older balloon data, and the narrow blue band by the new WMAP 
data. The dotted red vertical lines are different Higgs masses, and the 
current LEP bound produces the lower bound on $m_{1/2}$. The light blue region 
is excluded if $\delta a_{\mu} > 11 \times 10^{-10}$.(Other lines are discussed in text.)} 
\end{figure}

 \begin{figure}\vspace{-0cm}
\centerline{ \DESepsf(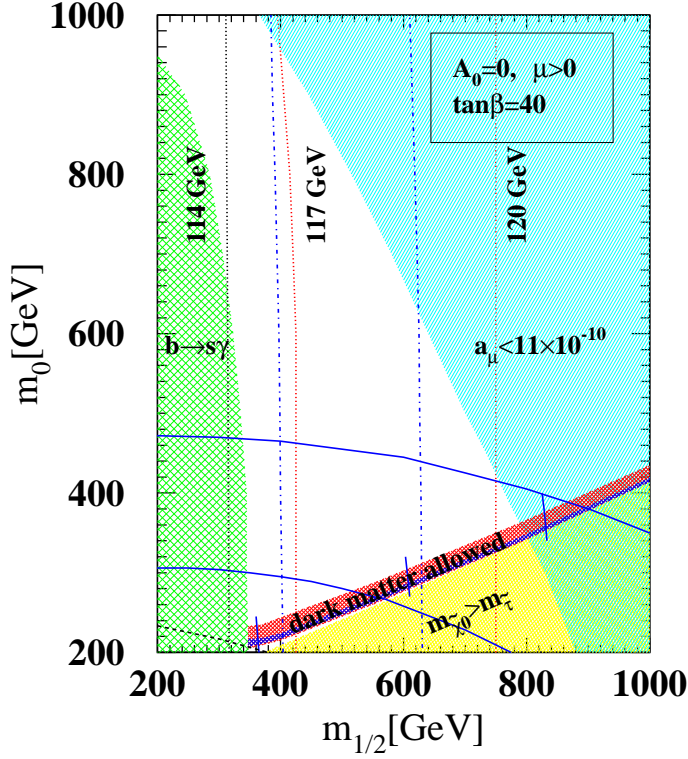 width 8 cm) }
\caption {\label{fig2}  Same as Fig. 1 for $\tan\beta = 40$, $A_0 =0$, $\mu >0$ except that now that 
the $b \rightarrow s  \gamma$ constraint (green region) produces the lower bound on
$m_{1/2}$.} 
\end{figure}

 \begin{figure}\vspace{-0cm}
\centerline{ \DESepsf(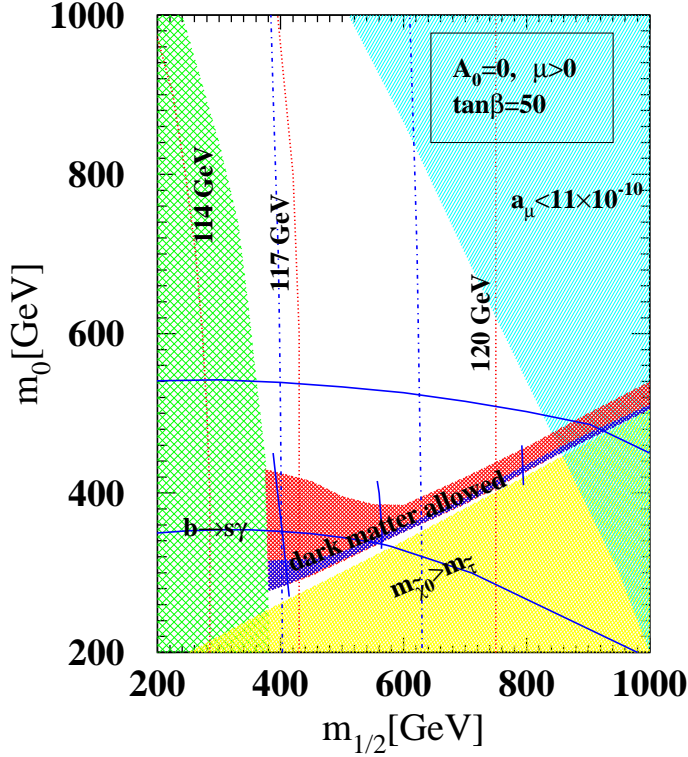 width 8 cm) }
\caption {\label{fig3}  Same as Fig. 2 for $\tan\beta = 50$, $A_0$ = 0, $\mu > 0$. Note that the 
large bulge at lower $m_{1/2}$ allowed by the older balloon data is now mostly 
excluded by the WMAP data.} 
\end{figure}

\section{Possible mSUGRA Signals At The 500-GeV LC}

To pair produce SUSY particles at a 500-GeV linear collider requires the 
sparticle mass to be less than 250 GeV. The previous constraints discussed, 
$m_{1/2} > 300$ GeV and the dark matter allowed bands, already excludes the pair 
production of the charginos, the heavier neutralinos, $\tilde\chi^0_{2,3,4}$, squarks, 
gluons and the heavy stau, $\tilde\tau_2$. Thus the remaining possible signals for 
mSUGRA are the following:
\begin{eqnarray}
            e^+ + e^- &\rightarrow& \tilde e + \tilde e \rightarrow  
	    (e^+ +\tilde\chi^0_1) +(e^-+\tilde\chi^0_1) \\ 
	    e^+ + e^- &\rightarrow& \tilde \mu + \tilde \mu \rightarrow  
	    (\mu^+ +\tilde\chi^0_1) +(\mu^-+\tilde\chi^0_1) \\ 
	    e^+ + e^- &\rightarrow& \tilde\tau_1^+ + \tilde\tau_1^- \rightarrow 
( \tau^++\tilde\chi^0_1) + ( \tau^- +\tilde\chi^0_1)\\
e^+ + e^- &\rightarrow& \tilde\chi^0_2 + \tilde\chi^0_1 
\rightarrow( \tau +\tilde\tau_1) + \tilde\chi^0_1\rightarrow  
	    (\tau^+ + \tau^- +\tilde\chi^0_1) +\tilde\chi^0_1 
\end{eqnarray}
Processes (4) and (5) are easiest to detect since the e and $\mu$ can be 
readily observed. Process (6) is more readily produced since L-R mixing in 
the tau~ mass matrix makes $m_{\tilde\tau_1} < m_{\tilde e,\,\tilde \mu}$, 
but hadronic tau 
identification (ID) is more difficult. In mSUGRA, 
$m_{\tilde\chi^0_2} \cong 2 m_{\tilde\chi^0_1}$ and 
so process (7) is kinematically feasible for a 500-GeV machine, and since 
the ${\tilde\tau_1}$ is the lightest slepton, the final state will be mainly taus as 
indicated. This allows us to divide the mSUGRA parameter space into three 
parts:\\

\noindent (1) $m_{\tilde e,\,\tilde\mu} < 250$ GeV. Here processes (4) and (5) can occur and SUSY can 
be detected with high precision. The kinematic reach of these processes is 
shown by the dashed black curve in the lower left hand corners of Figs. 1 
and 2. One sees that when other experimental constraints are taken into 
account, these processes require $\tan\beta < $40. Most of the previous 
analyses have been concentrated on this region of the parameter space. We note that $\tilde\tau_1$ pair
production can also occur in this parameter region.

\noindent (2) $m_{\tilde e,\,\tilde\mu} >$ 250 GeV, but $m_{\tilde\tau_1} <$ 250 GeV. Here processes (6) and (7) 
are the only ones possible, process (7) being possible if 
$m_{1/2}\stackrel{<}{\sim}400$ GeV. The kinematic reach for  process (6) is shown 
by the solid blue lines in Figs. 1-3, the lower one being for a 500-GeV 
machine, and the upper one for a 800-GeV machine. The kinematic reach of 
process (7) is shown in Figs. 1-3 as verticle blue dot-dash lines, the 
left hand one for a 500-GeV machine and the right hand one for an 800-GeV 
machine. We see that process (6) has a significantly larger reach than 
process (7), but neither cover much of the SUSY parameter space for a 500-GeV LC.

\noindent (3) $m_{\tilde e,\,\tilde mu,\,\tilde\tau_1} >250$ GeV and 
$m_{1/2}\stackrel{>}{\sim}400$ GeV. All final sparticle states 
are inaccessible and SUSY cannot be seen at a 500-GeV machine. 
(Unfortunately, this is not a small part of the parameter space allowed by 
mSUGRA.)

We consider first the region  where{ $m_{\tilde e,\,\tilde \mu} <$ 250 GeV}

This region of parameter space, where one can pair produce selectrons and 
smuons at a 500-GeV LC, occurs for low and intermediate tanbeta. As can be 
seen from Fig. 2, one requires $\tan\beta < 40$. Most of the $\tan\beta = 30$ 
parameter space is also inacessible, and one can see from Fig. 1, that the 
reach is not large even for $\tan\beta = 10$. There has been much study of this 
region (e.g. \cite{noj,tesla,jlc}) as it allows very accurate determination of SUSY 
masses. The basic reaction is
\begin{equation}
              e^+ + e^- \rightarrow \tilde e_R^+ + \tilde e_R^- \rightarrow  
	    (e^+ +\tilde\chi^0_1) +(e^-+\tilde\chi^0_1)\end{equation}
and the signal is thus $e^+ + e^- +$ \mpt (with a similar signal in the $\mu$ 
channel).

Since the processes are all two body, the kinematics is simple: there is a 
maximum and minimum lepton energy given by
\begin{equation}E_{min,max} ={m_{\tilde e_R}\over 2}[1-{m_{\tilde\chi^0_1}^2\over
{m_{\tilde e_R}^2}}]\gamma(1\mp\beta)\end{equation}
where $\gamma = \sqrt{s}/2m_{\tilde e_R}$ and $\beta = \sqrt{[1 -
4m_{\tilde\mu_R}^2/s]}$.  A measurement of $E_{min}$ and $E_{max}$ then 
determines $m_{\tilde e_R}$ and $m_{\tilde\chi^0_1}$ at the 
$1/10$\% level \cite{hm,pg}. In mSUGRA then, 
this would determine $m_0$ and $m_{1/2}$ very accurately.


Since the $\tilde\tau_1$ is the lightest slepton, pair production of these can also 
occur in this part of the parameter space.The corresponding tau~ analysis, 
however,  is more complicated for two reasons:\\ \noindent(1) there is L-R mixing in 
the $\tilde \tau$ mass matrix:
\begin{eqnarray}
m_{\tilde \tau}^2 ~=~\left(\begin{array}{cc} m_{LL}^2 &
m_{\tau}(A_{\tau}-\mu\tan\beta)\\
 m_{\tau}(A_{\tau}-\mu\tan\beta)&m_{RR}^2\end{array} \right)
\end{eqnarray}
with two eigenstates $\tilde\tau_1$and $\tilde\tau_2$
\begin{eqnarray}
\left(\begin{array}{c} \tilde\tau_1\\
\tilde\tau_2 \end{array}\right) ~=~\left(\begin{array}{cc} cos\theta_{\tau} &
sin\theta_{\tau}\\
-sin\theta_{\tau}&cos\theta_{\tau}\end{array} \right)
\left(\begin{array}{c} \tilde\tau_L\\
\tilde\tau_R\end{array} \right)
\end{eqnarray}
In mSUGRA the lightest stau, $\tilde\tau_1$, is the next to lightest SUSY particle 
(NLSP), and is mostly $\tilde\tau_R$.

\noindent(2) The decay pattern is now multiparticle i.e.:
\begin{equation}
e^+ + e^- \rightarrow \tilde\tau_1^+ + \tilde\tau_1^- \rightarrow 
( \tau^+ +\tilde\chi^0_1) + ( \tau^- +\tilde\chi^0_1)\rightarrow [jets + 
\bar{\nu_{\tau}} + X] + [jets + \nu_{\tau} + X]\end{equation}
where the jets are mainly $\pi$, $\rho$ and $a_1$ mesons i.e. the hadronic taus 
give rise to multipion states plus \mpt.

The analysis now is quite complicated \cite{noj} but using the known value of 
$m_{\tilde\chi^0_1}$ from the slectron and smuon analysis, the mass of the 
$\tau_1$ can be 
gotten at the $1/2$\% level \cite{hm,pg}. However, these analyses assume that $\delta 
m = m_{\tilde \tau_1} - m_{\tilde\chi^0_1}$ is (40 - 50) GeV. Such situations don't apply to 
mSUGRA, where due to the narrow co-annihilation band (see Figs. 1,2,3) one 
has $\delta m \cong$(5 - 15) GeV. The final state taus are thus much softer and 
so harder to identify and we discuss this case next.

In region (2) where  $m_{\tilde e_R,\,\tilde\mu_R} >$ 250 GeV the only slepton that can be pair
produced is the $\tilde\tau_1$.

This situation occurs for $\tan\beta >$ 40, and also for large parts of the 
parameter space with $\tan\beta <$ 40. For $\tan\beta >$ 40, the remaining possible signals of mSUGRA 
are then Eq. (6) and also Eq.(7) can occur for $m_{1/2}\stackrel{<}{\sim} 400$ GeV. Thus the signal in both cases are two 
$\tau's + $\mpt\ with acoplanar jets in the final state. As discussed in Sec. 2, the $\tilde\tau_1$ 
pair production has more reach as it extends to higher values of $m_{1/2}$, as 
can be seen in Figs. 1,2,3.  The narrowness of the co-annihilation band means that $\delta m$ is 
now quite small, and the techniques used earlier to detect SUSY signal with 
taus\cite{noj,tesla,jlc} no longer are applicable. To investigate what mass 
measurements might be made in this most difficult region of the parameter 
space, we have examined the three points of Table 1 which span the allowed 
dark matter band at $m_{1/2} = 360$ GeV, $\tan\beta = 40$ , $A_0 =0$, $\mu >$ 0.
\footnote{preliminary analysis of this case has been given in ref.\cite{teruki}}

\begin{table}[t]
\caption{Masses (in GeV) of SUSY particles for Points 1, 2 and 3. These 
points satisfy all the existing experimental bounds on mSUGRA.}\begin{center}
\begin{tabular}{l c c c c }
\hline \hline  MC &$m_{\tilde\chi^0_2}$ &  $m_{\tilde\tau_1}$ & 
$m_{\tilde\chi^0_1}$ &
        $\delta m$  \\
 Point & & & & $\equiv m_{\tilde\tau_1} -m_{\tilde\chi^0_1}$ \\
\hline  1. &   266 &  149.9 &  144.2
        &  5.7   \\
 2. &  266 &  154.8 &  144.2 &  10.6  \\
 3.  &  266 &  164.4 &  144.2 &  20.2 \\
\hline \hline
\end{tabular}
\end{center}
\end{table}
We use RH polarization $P(e^-) = -0.9$ to enhance the $\tilde\tau_1 \tilde\tau_1$ signal, 
and LH polarization, $P(e^-) = +0.9$ to see the $\tilde\chi^0_1 \tilde\chi^0_2$ signal. The 
production cross sections are given in Table 2. One sees that a significant 
signal exits provided the SM  four fermion background can be suppressed. 
These SM  backgrounds fall into two classes: (1) $\bar\nu \nu \tau^+ 
\tau^-$ sates arising from WW, ZZ etc. production, and (2) two photon 
processes $e^+ e^- \rightarrow \gamma^* \gamma^* +  e^+ e^-\rightarrow \tau^+ \tau^-$ 
(or $q \bar{q}$) + $e^+ + e^-$ 
where the final state $e^+ e^-$ pair are at a  small angle to the beam pipe and 
the $q\bar{q}$ jets fake a $\tau^+ \tau^-$ pair. The LH polarization cuts are 
optomized to enhance the $\tilde\chi^0_1 \tilde\chi^0_2$ signal and the RH to optomize the the 
$\tilde\tau_1\tilde\tau_1$ signal. The cuts chosen were the following:

\noindent(1) 2 $\tau '$s:
$N_{jet} = 2$ with $E_{jet}>3$ GeV for $Y_{cut}\geq 0.0025$\cite{jade}, 
each  passing the $\tau_h$ ID of 1 or 3 tracks 
 and the two jets are
oppositely charged ($q_1 q_2 = -1$).

\noindent(2) To reduce the WW, ZZ, Z-$\gamma^*$, etc. background,
Acoplanarity $> 40^o$ along with\\\noindent
LH Polarization: $-q_{jet} cos(\theta_{jet}) < 0.7$; $-0.8 < cos[\theta(j_2,P_{vis})] 
< 0.7$\\\noindent
RH Polarization: $|cos(\theta_{jet})| < 0.65$; $-0.6 < cos[\theta(j_2,P_{vis})] <
0.6$.

\noindent(3) To reduce the 2 photon ($\gamma^* \gamma^*$) events:
Veto EM cluster in $5^o < \theta < 28^o$ with $E > 2$ GeV;
Veto electrons within $\theta > 28^o$ with $p_T > 1.5$ GeV; 
Veto EM clusters ($e^+/e^-$) 
in $2^o (1^o) < \theta_{cluster} < 5.8^o$  with $E_{cluster} > 100$ GeV 
using an active  beam mask of $2^o (1^o)$ - $5.8^o$.

\noindent(4) We examine \mpt$ > 5$, 10 or 20 GeV.

The Monte Carlo analysis was done using the following programs: 
\noindent(1) ISAJET 
7.63 to generate SUSY events. (2) WPHACT v2.02 pol for SM backgrounds. (3) 
Tauola v2.6 for tau decay. (4) Events 
were simulated and analysed with LCD 
Root Package  v3.5 with LD Mar 01 detector parametrization.
\begin{table}[t]
\caption{SUSY and SM production cross sections (in fb)  for 
polarizations  $P(e^-) = -0.9$(RH), 0, and +0.9.}
\begin{center} 
\begin{tabular}{l| c| c| c }
\hline \hline $P$($e^-$)&-0.9(RH)&0&0.9(LH)\\\hline
SM&7.84&48.9&89.8\\
SUSY&$\tilde\chi^0_2\tilde\chi^0_1$, $\tilde\tau_1\tilde\tau_1$&&\\\hline
1.& 0.53,  26.4&3.39,  19.6&7.10,  12.8\\
2. &0.52,  24.4&3.31,  18.4&6.91,  11.8\\
3. &0.50,  21.1&3.15,  15.8&6.62,  10.3\\
\hline \hline
\end{tabular}\end{center}\end{table}

The number of events for each class of final states for the case \mpt$> 5$, 10, 20 
GeV is given in Table 3, and the significance ($N_s$/$\sqrt{N_B}$) for \mpt$ > 5$, 10, 20 GeV is shown in Table 4. (For the RH 
polarization, the $\tilde\chi^0_1 \tilde\chi^0_2$ events are treated as background and for the LH 
polarization the $\tilde\tau_1 \tilde\tau_1$ events are treated as background.) One sees 
that the RH polarization strongly suppresses the WW etc. SM background and 
the neutralino events, and combined with a $2^o$ mask it leaves a clean 
signal for the $\tilde\tau_1 \tilde\tau_1$ events. The LH polarization then allows for 
the detection of the $\tilde\chi^0_1 \tilde\chi^0_2$ signal. With no mask there would be $\sim20,000$ 
additional SM background events swamping the SUSY signal. Thus the mask is 
essential to detect SUSY in this region of parameter space.

 From Table 4 one sees that one has a robust discovery significance for 
both SUSY signals, the $\tilde\chi^0_1 \tilde\chi^0_2$ case for all $\delta m$ and all minimum \mpt, 
and the $\tilde\tau_1 \tilde\tau_1$ for all  $\delta m$ with minimum \mpt$= 5$ GeV.  Using the 
above acceptances we can find the 5 $\sigma$ (discovery) reach for each 
signal. For \mpt$>$ 5 GeV we find for $\delta m = 5$ GeV that 
$m_{\tilde\chi^0_2} = 286$ GeV 
(corresponding to $m_{1/2} = 383$ GeV) and for $\delta m = 20$ GeV that 
$m_{\tilde\chi^0_2} = 295$ 
GeV ($m_{1/2} =396$ GeV). The $5 \sigma$ reach is larger for the $\tilde\tau_1$. We find 
for $\delta m = 5$ GeV that   $m_{\tilde\tau_1} = 232$ GeV (corresponding to $m_{1/2} = $514 
GeV ) and for $\delta m = 20$ GeV that $m_{\tilde\tau_1} = 193$ GeV ($m_{1/2}  = 463$ GeV).

\begin{table}[t]\caption{Number of events for luminosity of 500 fb$^{-1}$  for 
points 1, 2 and 3 with $\delta m = 5$, 10, 20 GeV.}
\begin{center}
\begin{tabular}{l lr | rrr | rrr}
\hline
\hline
  & & &         \multicolumn{3}{c|}{${\cal P}(e^-)=0.9$(L.H.)} &
        \multicolumn{3}{c}{${\cal P}(e^-)=-0.9$(R.H.)} \\
Process &   &  & $\mpt^{min}$ = 5  & 10 & 20 & 5 & 10 & 20 \\
\hline
$\tilde\chi^0_1\tilde\chi^0_2$
        & Pt.1 & & { 549} & { 495} & { 367} &
                        26 & 24 & 17 \\
        & Pt.2 & & {\ 777} & { 714} & { 518} &
                        33 & 31 & 22 \\
        & Pt.3 & & { 886} & { 831} & { 622} &
                        36 & 34 & 24 \\
\hline
$\tilde\tau_1\tilde\tau_1$
        & Pt.1 & & 151 & 16 & 0 &
                        { 241} & 30 & 0\\
        & Pt.2 & & 584 & 344 & 25 &
                        { 811} & { 500} & 37 \\
        & Pt.3 & & 935 & 781 & 356 &
                        { 1244} & { 1074} & { 526} \\
\hline
SM weak    &  &    & 1745 & 1626 & 1241 & 129 & 123 & 100 \\
$e^+e^-\tau^+ \tau^-$ & ~$2^\circ-5.8^\circ$ mask & & 449 & 5 & 0 & 210 & 2 & 0\\
             & [$1^\circ-5.8^\circ$ mask] &  & [4] & [0] & [0] & [2] & [0] & [0]\\
$e^+e^-q \bar{q}$ & ~$2^\circ-5.8^\circ$ mask & & 79 & 1 & 0 & 38 & 1 & 0\\
             & [$1^\circ-5.8^\circ$ mask] &  & [1] & [0] & [0] & [0] & [0] & [0]\\
\hline
\hline
\end{tabular}
\end{center}
\label{tab:LC500_Results}
\end{table}
\begin{table}[t]\caption{Significance ($N_S/ {\sqrt{N_B}}$) for a luminosity of 500 fb$^{-1}$ for 
SUSY discovery for points 1, 2, and 3.}
\begin{center}
\begin{tabular}{|c| c| c c c|  }
\hline \hline  Min. $\mpt$& &5&10&20\\\hline
Proc.&$\delta m$&&&\\\hline
$\tilde\chi^0_1\tilde\chi^0_2$(LH)&5&11.2&12.2&10.4\\
&10&14.5&16.1&14.6\\
&20&15.6&17.0&14.5\\\hline\hline
$\tilde\tau_1\tilde\tau_1$(RH)&5&12&2.5&0\\
&10&40&40&3.3\\
&20&61&85.6&47\\\hline\hline
\end{tabular}
\end{center}\end{table}

\section{Conclusions}

To examine what mSUGRA signals could be observed at a 500-GeV linear 
collider, it is necessary to take account of all the existing constraints 
on the parameter space. Particularly important are the light Higgs mass, 
$b \rightarrow s \gamma$ and the dark matter constraints. The former two require 
$m_{1/2}\stackrel{>}{\sim} 300$ GeV, and the latter  for the domain 
$m_0, \,m_{1/2} \stackrel{<}{\sim}$ 1 TeV require 
$\delta m = m_{\tilde\tau_1} - m_{\tilde\chi^0_1} \stackrel{<}{\sim}$ 15 GeV due to the narrowness of the 
co-annihilation band. Possible signals at a 500-GeV LC then fall into two 
types:

(1) If $m_{\tilde e,\,\tilde\mu}<$ 250 GeV, then $\tilde e$ and $\tilde\mu$ pair production is possible, and 
the selectron and smuon masses can be measured with very high accuracy. 
This region covers less and less of the parameter space as $\tan\beta$ 
increases, and has an upper bound of $\tan\beta $= 40.

(2) In general the lightest stau is the lightest slepton, and there is a 
parameter region where pair production of the $\tilde\tau_1$ can occur over the 
full $\tan\beta$ domain. In addition, the production of $\tilde\chi^0_1+\tilde\chi^0_2$ is 
possible for $m_{1/2}\stackrel{<}{\sim}$ 400 GeV. The analysis of the stau pair production is 
greatly complicated by the smallness of $\delta m$. To examine the 
difficulties, we have considered here the case where $m_{1/2}$= 360 GeV , 
$\tan\beta = 40$, $\mu >0$. Signals of both these processes can be detected for 
$\delta m > 5$ GeV provided an active mask down to $2^o$ can be constructed to 
eliminate the $\gamma^{\*}\gamma^{\*}$ processes (where final states $e^+ e^- \tau^+ \tau^-$ 
(or $q \bar q$ faking a $\tau$ pair) with the $e^+$ and $e^-$ at very small angles to 
the beam pipe) occur. Polarized beams are also needed, RH polarization to 
suppress SM processes and enhance the $\tilde\tau_1\tilde\tau_1$ process, and LH to see 
the $\tilde\chi^0_1\tilde\chi^0_2$ process. We find that the 5$\sigma$ discovery reach for 
$\delta m >$ 5 GeV is $m_{\tilde\tau_1}$ = 232 GeV (corresponding to $m_{1/2}$= 514 GeV) 
and $m_{\tilde\chi^0_2}$=286 GeV(corresponds to $m_{1/2}$= 383 GeV). We are analysing how accurately  one might determine the $\tilde\tau_1$ and neutralino 
masses. (For a discussion of how this might be done see \cite{bd}.)

(3) The above signals cover perhaps less that one half of the parameter 
space assuming the current $g_\mu - 2$ bound maintains, and even less if this 
bound shrinks. Thus over a large amount of the parameter parameter space, a 
500-GeV machine would see no signal of SUSY if mSUGRA (or a SUSY theory 
like it) were valid. This strongly argues for building an 800-GeV linear 
collider, where the kinematic regions where SUSY can be seen increases 
significantly. The discussion of SUSY detection reach for an 800-GeV machine 
would require, however, a separate analysis of the type discussed here, and 
detection of the $\tilde\tau_1$ pair production in the co-annihilation region would 
likely require an active mask down to $1^o$. 
\section{Acknowledgements}
This work is supported in part by the National Science Foundation Grant
PHY-0101015 and in part by  the Natural Sciences and Engineering Research Council 
of Canada and by the Department of Energy Grant DE-FG03-95ER40917.


\begin{thebibliography}{99}
\bibitem{sugra1}A.H. Chamseddine, R. Arnowitt, and P. Nath,
\Journal{\PRL}{49}{1982}{970}. 
\bibitem{sugra2}R.~Barbieri, S.~Ferrara, and C.~A.~Savoy,
\Journal{\PRD}{119}{1982}{343}; L. Hall, J. Lykken, and S. Weinberg,
\Journal{\PRD}{27}{1983}{2359}; P. Nath, R. Arnowitt, and A.H. Chamseddine,
\Journal{\NPB}{227}{1983}{121}.
\bibitem{nilles}For a review, P. Nilles, {\it Phys. Rept.} {\bf 110}, {1984} (1).
\bibitem{higgs1} P.~Igo-Kemenes, LEPC meeting,   
(http://lephiggs.web.cern. ch/LEPHIGGS/talks/index.html).
\bibitem{bsgamma} M. Alam et al., \Journal{\PRL}{74}{1995}{2885}. 
\bibitem{sp} D.N Spergel et al., astro-ph/0302209.
\bibitem{aleph}The ALEPH collaboration, ALEPH-CONF.2001-009.
\bibitem{BNL}G.Bennett et.al., Muon (g-2) Collaboration, 
\Journal{\PRL}{89}{2002}{101804}.
\bibitem{dav}M. Davier, S. Eidelman, A. Hocker, and Z. Zhang,   {\it Eur. Phys.
J} {\bf C27} 2003 (497).
\bibitem{hag}K. Hagiwara, A. Martin, D. Nomura, and T. Teubner,  
\Journal{\PLB}{557}{2003}{69}.
\bibitem{jeg}F. Jegerlehner, {\it J. Phys.}, {\bf G29}, {2003} (101).
\bibitem{akh}R.R. Akhmetshin et al., hep-ex/0308008.
\bibitem{ellis}J. Ellis, K. Olive, Y. Santoso, and V. Spanos, hep-ph/0303043.
\bibitem{rattazi}R. Rattazi, and U. Sarid, \Journal{\PRD}{53}{1996}{1553};
 M. Carena, M. Olechowski, S. Pokorski, and C. Wagner,\Journal{\NPB}{426}{1994}{269}.
\bibitem{degrassi}G. Degrassi, P. Gambino, and G. Giudice, 
\Journal{\it JHEP} {\bf 0012}, {2000} {(009)}; M. Carena, D. Garcia, U. Nierste,
and C. Wagner,
\Journal{\PLB} {499}{2001}{141}, D'Ambrosio, G. Giudice, G. Isidori, and A.Strumia, hep-ph/0207036;
A. Buras, P. Chankowski, J. Rosiek, and L. Slawianowska, hep-ph/0210145.
\bibitem{bdutta} R. Arnowitt, B. Dutta, and Y. Santoso, hep-ph/0010244; 
hep-ph/0101020; \Journal{\NPB}{606}{2001}{59}; 
J Ellis, T. Falk, G. Ganis, K. Olive, and M. Srednicki,
\Journal{\PLB} {570}{2001}{236};  
J. Ellis, T. Falk, K. Olive,\Journal{\PLB}{444}{1998}{367};
J. Ellis, T. Falk, K. Olive, and M. Srednicki, {\it AP} {\bf 13},{2000} {(181)};
Erratum-ibid.{ \bf 15} {2001} {413}; M. Gomez, and J. Vergados, \Journal{\PLB}{512}{2001}{252}; 
M. Gomez, G. Lazarides, C.
Pallis, \Journal{\PRD}{61}{2000}{123512}; \Journal{\PLB}{487}{2000}{313};
 L. Roszkowski, R. Austri, and T. Nihei, {\it JHEP} {\bf 0108}, {2001} {(024)}; 
 A. Lahanas, D. Nanopoulos, and V. Spanos, \Journal{\PLB}{518}{2001}{518}.
\bibitem{noj}M. Nojiri, K. Fujii, and T. Tsukamoto, \Journal{\PRD}{54}{1996}{6756}.
 \bibitem{tesla}TESLA TDR, DESY 2001-011.
\bibitem{jlc}JLC, hep-ph/0109166.
\bibitem{hm}H-U. Martyn, ECFA/DESY Workshop, prague, november, 2002,
www-hep2.fzu.cz/ecfadesy/ECFA$_{-}${DESY}$_{-}${Praha2002.htm}.
\bibitem{pg}P. Granis, Proceedings of 2002 International Workshop on Linear Colliders, Jeju Island, Korea,
hep-ex/0211002.
\bibitem{teruki}T. Kamon, R. Arnowitt, B. Dutta, and V. Khotilovich, Proceedings of 2002 International Workshop on Linear Colliders, Jeju Island, Korea,
hep-ph/0302249.
\bibitem{jade}JADE collaboration, W. Bartl et al., {\it Z. Phys} {\bf C33}, {1986} {(23)};
 S. Bethke et al., \Journal{\PLB}{213}{1988}{235}.
\bibitem{bd}B. Dutta, R. Arnowitt,  T. Kamon, and V. Khotilovich, 2003 American LC Workshop, Cornell University, 
Ithaca, NY, www.lns.cornell.edu/public/LC/workshop/workinggroups.html

\end{thebibliography}
\end{document}